\documentclass[aip,apl,numerical,twocolumn,reprint,footinbib]{revtex4-1}

\usepackage[latin1]{inputenc}
\usepackage[english]{babel}
\usepackage{graphpap}
\usepackage[dvips]{graphicx}
\usepackage{dcolumn}
\usepackage{xspace}
\usepackage{ifthen}		
\usepackage[usenames,dvipsnames]{color}
\usepackage{amsfonts,amsmath,amssymb}	
\usepackage[tight]{units}

\begin{document}
\newcommand{\todo}[1]{\textcolor{blue}{#1}}
\newcommand{\hl}[1]{\textcolor{red}{#1}}
\newcommand{\Fig}[1]{Fig.~\ref{#1}\xspace}
\newcommand{\Eq}[1]{Eq.~(\ref{#1})\xspace}
\newcommand{\eq}[1]{(\ref{#1})\xspace}
\renewcommand{\vec}[1]{\mathbf{#1}}
\renewcommand{\i}{\ensuremath{{\rm i}\,}}
\newcommand{\half}{\ensuremath{ \nicefrac{1}{2}\;}}
\newcommand{\abs}[1]{\left|#1\right|}	
\newcommand{\diff}[3]{\ifthenelse	{\equal{#3}{1}}
								{\frac{{\rm d} #1}{{\rm d} #2}}
								{\frac{{\rm d}^{#3} #1}{{\rm d} #2^{#3}}}	}
\newcommand{\diffop}[2]{\ifthenelse	{\equal{#2}{1}}
								{\frac{{\rm d}}{{\rm d} #1}}
								{\frac{{\rm d}^{#2}}{{\rm d} #1^{#2}}}	}
\newcommand{\ket}[1]{\left| #1 \right\rangle}
\newcommand{\bracket}[3]{\left\langle #1 \right| #3 \left| #2 \right\rangle}	
\newcommand{\op}[1]{\hat{\rm #1}}														
\newcommand{\vecop}[1]{\hat{\vec{#1}}}
\newcommand{\muB}{\ensuremath{\mu_{\rm B}}}		
\newcommand{\muorb}{\ensuremath{\mu_{\rm orb}}}	
\newcommand{\muS}{\ensuremath{\mu_{\rm s}}}		
\newcommand{\DKK}{\ensuremath{\Delta_{K K'}}}		
\newcommand{\Dso}{\ensuremath{\Delta_{\rm so}}}	
\newcommand{\OmegaR}{\ensuremath{\Omega_{\rm R}}}	
\title{Readout of carbon nanotube vibrations based on spin-phonon coupling}

\author{C. Ohm}
\address{Institut f\"ur Theorie der Statistischen Physik, RWTH Aachen University, 52056 Aachen, Germany}
\address{JARA -- Fundamentals of Future Information Technology}

\author{C. Stampfer}
\address{JARA -- Fundamentals of Future Information Technology}
\address{II. Physikalisches Institut B, RWTH Aachen University, 52056 Aachen, Germany}
\address{Peter Gr\"unberg Institut, Forschungszentrum  J\"ulich, 52425 J\"ulich, Germany}

\author{J. Splettstoesser} 
\address{Institut f\"ur Theorie der Statistischen Physik, RWTH Aachen University, 52056 Aachen, Germany}
\address{JARA -- Fundamentals of Future Information Technology}

\author{M.~R. Wegewijs}
\address{Institut f\"ur Theorie der Statistischen Physik, RWTH Aachen University, 52056 Aachen, Germany}
\address{JARA -- Fundamentals of Future Information Technology}
\address{Peter Gr\"unberg Institut, Forschungszentrum  J\"ulich, 52425 J\"ulich, Germany}

\email{ohm@physik.rwth-aachen.de}
\date{\today}

\begin{abstract}
We propose a scheme for spin-based detection of the bending motion in suspended carbon-nanotubes, using the curvature-induced spin-orbit interaction. We show that the resulting effective spin-phonon coupling can be used to down-convert the high-frequency vibration-modulated spin-orbit field to spin-flip processes at a much lower frequency. This vibration-induced spin-resonance can be controlled with an axial magnetic field. We propose a Pauli spin blockade readout scheme and predict that the leakage current shows pronounced peaks as a function of the external magnetic field. Whereas the resonant peaks allow for frequency readout, the slightly off-resonant current is sensitive to the vibration amplitude.
\end{abstract}

\maketitle 

Carbon nanotubes (CNTs) have been shown to be an ideal playground for realizing both spintronic nanodevices~\cite{Cottet:2006uq,Kuemmeth:2011kx} and high-frequency nano-electromechanical systems with high-Q resonators.\cite{Sapmaz:2003fk,Sazonova:2004vn,
Garcia-Sanchez:2007ul,Huttel:2009fk,Lassagne:2009kx,Steele:2009ve} This makes nanotubes interesting candidates for the development of quantum electromechanical systems,\cite{Schwab:2005ys,Zippilli:2009qy} relying however crucially on advanced schemes for sensitive motion detection and electromechanical readout. In this Letter, we propose a new scheme of built-in, spin-based motion detection, which allows for a sensitive readout of ultra-high frequencies and tiny vibration amplitudes. 
Indeed, the spin-orbit interaction (SOI), recently demonstrated to be important in CNTs, \cite{Kuemmeth:2008fk, Jespersen:2010fk} provides a new way to weakly couple the electron spin to the vibrational motion through the effect of the motion on electron-orbits.
To utilize the cross talk of spin and vibrations via SOI for the readout of the vibration frequency and amplitude, one needs a selective sensitivity to spin-flip processes. This is provided by the well established technique of Pauli spin blockade spectroscopy, which was successfully used to study the effects of nuclear and impurity spins in nanotube quantum dots.\cite{Buitelaar:2008lr,Churchill:2009fr,Churchill:2009mz,Chorley:2011ly}
The SOI breaks the fourfold degeneracy of the energy spectrum of the CNT,\cite{Kuemmeth:2008fk, Jespersen:2010fk} defining new effective Kramers two-level systems that have been the subject of valley-tronics proposals~\cite{Palyi:kx}, which have motivated this work.
In contrast, we here investigate a working point where the \textit{real} spin is the degree of freedom. In this regime, we propose a high-frequency vibration readout using the SOI induced spin-flip leakage current in a half-suspended double quantum dot. Most interestingly, this scheme provides a built-in down-mixing of the vibrational frequency, which may even enable time-resolved measurements of the nanotube vibrations. Moreover, it provides sensitive access to the vibrational amplitude. The scheme can additionally be considered as driving-free, a feature which has been emphazised to be important in Ref.~\onlinecite{Huettel:2010}. The outline of this Letter is as follows: We first discuss the central idea of down-mixing an effective high-frequency spin-orbit field, driven by the flexural nanotube vibrations, to a low-frequency modulation of a spin-flip rate. We then outline the derivation of this spin-orbit field from a microscopic  model. Finally, we describe a readout scheme for the vibrational frequency using the Pauli spin blockade.
\begin{figure}[t]
  \includegraphics[width=0.45\textwidth]{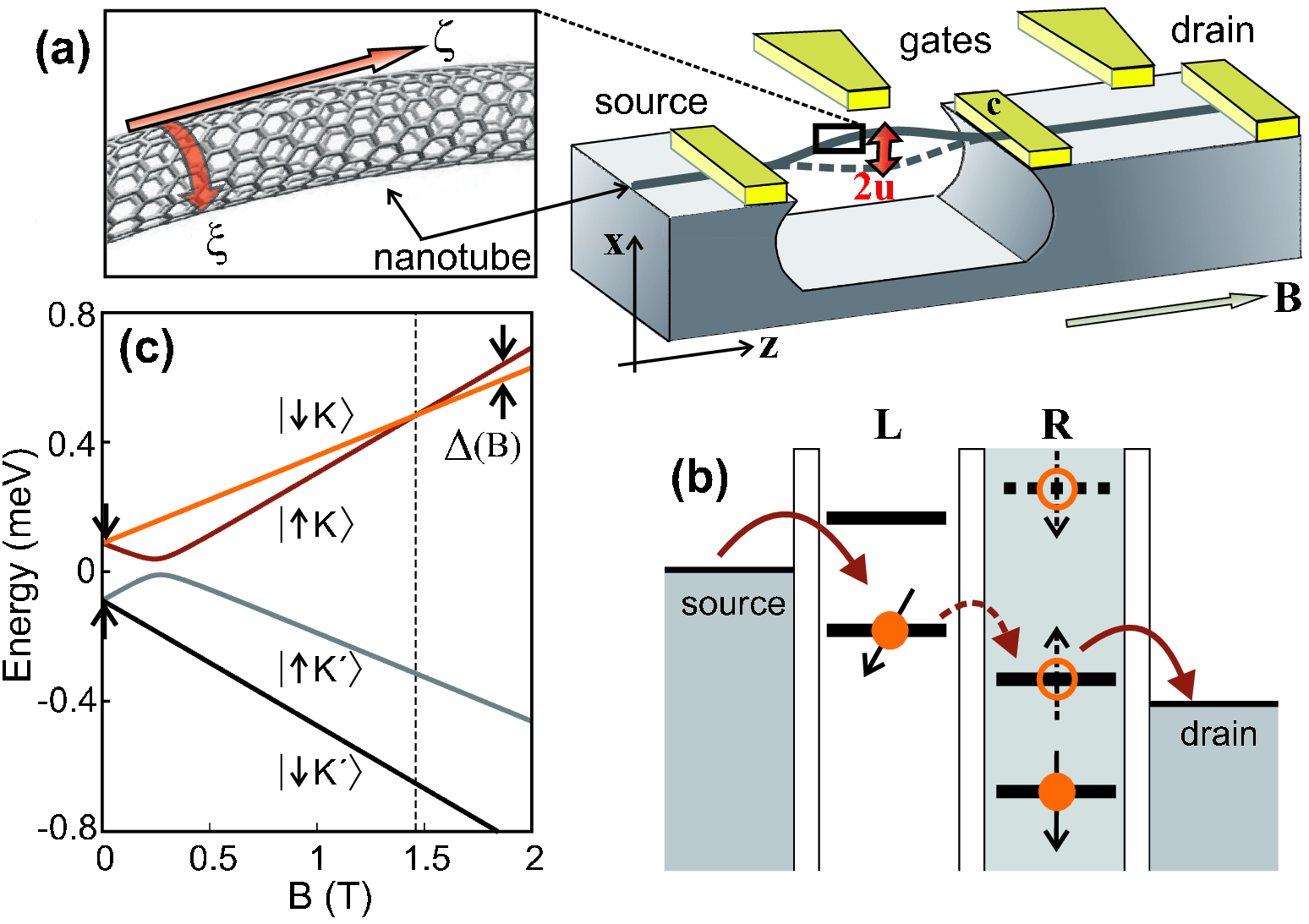}
  \caption{
    (a)	Setup with a partly suspended, vibrating CNT.
    (b) Schematic diagram of the Pauli spin blockade.
      Interdot tunneling is suppressed when an electron entering the left dot
      creates a triplet state with the electron trapped in the right one,
      since the triplet with two electrons on the right dot involves a high orbital excitation indicated by the dashed level.
      Vibrations of the left dot can lift this blockade by rotating the spin through spin-phonon coupling, creating a singlet state which allows interdot tunneling to proceed through the ground state orbital.
  	(c) Spectrum of a CNT, \Eq{Eq:Hamiltonian2}, as function of the magnetic field $B$ applied along the  \emph{unbent} CNT.
    }
  \label{Fig:setup}
\end{figure}

\emph{Down-mixing the high-frequency spin-orbit field.}
We consider a CNT, clamped between a source, a central gate and a drain electrode, see~\Fig{Fig:setup}(a). Only one part of the CNT is suspended while the other is resting on the substrate.  By electrostatic gating, two quantum dots are defined in the nanotube - one on the suspended part and one on the non-suspended part.  In this section we focus on the suspended, vibrating part of the nanotube.  As explained below, in the presence of a high magnetic field, it constitutes a well-defined two-level spin system (TLS) with an effective spin-phonon coupling due to the interplay between SOI and flexural vibrations of the tube.
The energy splitting $\Delta$ of the TLS  is tunable with the magnetic field. The flexural mode of the vibrating CNT, affecting the spin states via the spin-phonon coupling, is supposed to have a frequency  $\omega \approx \unit[5]{GHz} \sim \unit[3.3]{\mu eV}$.
\footnote{
The vibrational amplitude is bounded from below by the ground state vibrational fluctuations. Using Eq. (6) of Ref.~\onlinecite{Rudner:2010fj}, we find
  $\mathrm{d}u_x/\mathrm{d}z \sim q \sqrt{\hbar/(2M \omega(q))} \sim \sqrt{\hbar/(2M\beta)}$
using the dispersion relation $\omega(q)=\beta q^2$ with $\beta=\unit[1.28\times 10^{-4}]{m^2/s}$ and $M=2\pi R L \rho$ with mass density $\rho=\unit[9.66 \times 10^{-7}]{kg/m^2}$. Assuming a CNT length of $L=\unit[160]{nm}$ and diameter $d=\unit[2.0]{nm}$ we obtain $\omega(\pi/L) \approx \unit[5]{GHz}$ and with \Eq{eq:lambda}, $\lambda\approx \unit[0.016]{\mu eV}$.
}
The coupling constant, $\lambda$, depends on the vibrational amplitude as well as on the spin-orbit coupling strength.\cite{Rudner:2010fj}  A lower bound to the coupling strength, in the limit of zero-point fluctuations, is estimated to be $\lambda\approx \unit[0.016]{\mu eV}$.\cite{Note1} We consider that deflections are generally significantly larger than this. Hence the vibration can be treated classically, resulting in an effective time-dependent spin-orbit field. Assuming furthermore that the deflections are small w.r.t. the nanotube length, we obtain an electron spin resonance (ESR) type model:
\begin{equation}
  \op{H}_{\rm ESR} = \Delta \, \op{S}_z + \lambda \cos(\omega t) \, \op{S}_x \label{Eq:Hamiltonian1}
  ,
\end{equation}
where $\op{S}_x,\op{S}_z$ are the electron spin 1/2 operators. Importantly, when tuning $\Delta$ to match the frequency $\omega$, the coupling is small with respect to this, $\lambda\ll\Delta$, such that the narrow spin-resonance of width $\sim \lambda$ is contained in the range $\abs{\hbar\omega-\Delta} \ll \abs{\hbar\omega + \Delta}$, where the rotating wave approximation (RWA) is applicable. We then find that the transverse component of the spin performs a Larmor precession with a high-frequency $\Delta \sim \omega$, whereas the $z$ component oscillates at the Rabi-frequency $\OmegaR = \hbar^{-1} [ (\Delta - \hbar\omega)^2 + (\frac{\lambda}{2})^2 ]^{\half}$ and the time-dependent spin-flip probability is $W_{\uparrow \downarrow}(t) = \sin(\OmegaR t /2)\times \lambda^2 / (\lambda^2 + 4(\Delta-\hbar\omega)^2) $.
We observe:
(i) At resonance the spin-flip rate is maximal and the vibration frequency can be found from the resonant value of $\Delta=\hbar\omega$.
(ii) Slightly away from resonance, $\lambda \ll \abs{\Delta-\hbar\omega} \ll \abs{\Delta+ \hbar\omega}$, the spin-flip rate depends on the vibration amplitude as $\lambda^2 /(4(\Delta-\hbar\omega)^2)$ allowing $\lambda$ to be read out. 
(iii) At resonance the down-conversion of the original driving frequency is maximal, $\OmegaR \approx \lambda / 2\hbar \ll \omega$.
One can thus detect high-frequency vibrations by using the spin-phonon resonance effect introduced here, which is intrinsic to the CNT and which can be addressed by tuning a magnetic field. Inserting typical values for CNTs,
\footnote{
Typical parameters are $\Dso=\unit[0.17]{meV}$, $\DKK=\unit[0.05]{meV}$, $\muS=2 \;\muB$, $\muorb=5.7 \;\muB$.
} 
a down-mixing factor of ${\omega}/{\OmegaR} \sim 2 \hbar\omega/\lambda \approx 2400$ is obtained.
We emphasize that the weakness of the spin-phonon coupling $\lambda$ guarantees that both the width of the spin-flip resonance peak is small and that the down-mixing ratio is large.

\emph{Spin-phonon coupling in nanotubes.}
We now outline the derivation of the model (\ref{Eq:Hamiltonian1}).
In the presence of a strong confinement potential, the ground state multiplet of a CNT quantum dot is nearly  fourfold degenerate, and described by the product states $\ket{s \tau}$ of spin ($s\in\{\uparrow,\downarrow\}$) and isospin ($\tau\in\{K,K'\}$). Exact degeneracy is broken by spin-orbit interaction,~\cite{Kuemmeth:2008fk, Jespersen:2010fk} and by lattice disorder causing isospin mixing.~\cite{Palyi:kx} The quantum dot Hamiltonian for this subspace was derived in Ref.~\onlinecite{Flensberg:2010uq}:
\begin{align}
  \op{H}_{\rm d} =	 - \Dso\, \vec{e}_\zeta\cdot\vecop{S} \otimes \op{\tau}_3 - \frac{1}{2} \DKK \op{\tau}_1 \nonumber  \\
 					 + \muS\, \vec{B}\cdot\vecop{S}
  					 + \muorb\, \vec{B}\cdot\vec{e}_\zeta  \; \op{\tau}_3.
  \label{Eq:Hamiltonian2}
\end{align}
Here the curvature enhanced spin-orbit coupling, $\Dso$, leads to a spin polarization along the nanotube axis $\vec{e}_\zeta$.  The inter-valley scattering induced by lattice impurities and characterized by  $\DKK$, hybridizes the two valley (isospin) degrees of freedom. Additionally, the external magnetic field, $\vec{B}=B \vec{e}_{z}$, applied along the axis of the unbent CNT, gives rise to an orbital and spin Zeeman effect through $\muS$ and $\muorb$. For deflections with curvature radii much larger than the quantum dot length, the model \eq{Eq:Hamiltonian2} still applies.\cite{Flensberg:2010uq} The evolution of the spectrum with the magnetic field is shown in \Fig{Fig:setup}(c). At high fields the spectrum splits into two sectors with fixed isospin and we concentrate on the TLS defined in the isospin $K$ sector. The level-splitting between these two upper states $\ket{\uparrow K}$, $\ket{\downarrow K}$ is  given by 
$\Delta= \muS B +\half ( {[(\Dso - 2\muorb B )^2 + \DKK^2]^{\half}} -{[(\Dso + 2\muorb B)^2 + \DKK^2]^{\half} })$ with an exact crossing at 
$\abs{B_\mathrm{cross}} = \left[\nicefrac{\Dso^2}{\muS^2} - \nicefrac{\DKK^2}{(\muorb^2 - \muS^2)}\right]^{\half}$, 
indicated by the vertical line in \Fig{Fig:setup}(c).
The spin-phonon coupling can now be obtained by treating the vibrations as instantaneous deflections of the CNT.\cite{Rudner:2010fj} The displacement of a point on the nanotube is given by the displacement vector $\vec{u}(z,t)$, see \Fig{Fig:setup}(a). The time-dependent tangent vector of the vibrating CNT written as function of $z$, with the coordinate $z$ along the axis of the unbent nanotube, reads $\vec{e}_\zeta \approx \vec{e}_z + \diffop{z}{1} \vec{u}(z,t)$, where we consider deflections small compared to the nanotube length.
We assume that the bending vibration takes place in the $x$-direction and have a harmonic time-dependence. The resulting effective spin-phonon Hamiltonian is
$  \op{H}_{\rm s-ph} = \lambda \cos(\omega t) \, \op{S}_x \otimes \op{\tau}_3, $
with the coupling parameter
\begin{equation}
  \lambda=\Dso \diff{u_x}{z}{1}
  .
  \label{eq:lambda}
\end{equation}
The spectrum of a quantum dot located on the vibrating part of the CNT is consequently determined by  $\op{H}_{\rm d} + \op{H}_{\rm s-ph}$. For high magnetic fields $B \approx B_\mathrm{cross}$ the isospin decouples from the real spin states and we can project onto the isospin-$K$ sector, obtaining the ESR model \eq{Eq:Hamiltonian1}.

\begin{figure}[t]
  \includegraphics[width=0.40\textwidth]{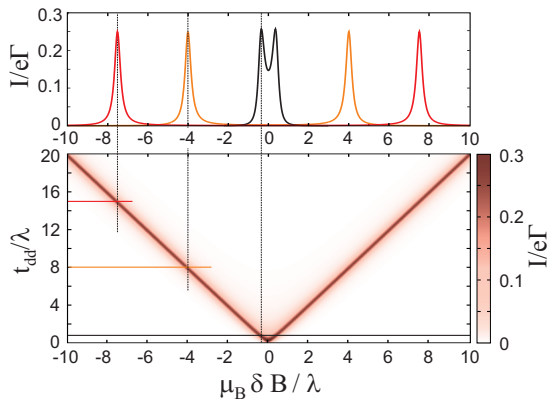}
  \caption{
    Charge current as function of
    the magnetic field detuning $\delta B = B - B_{\rm res}$ from the resonant value ($\Delta(B_{\rm res})=\omega$)
    and the inter-dot tunneling $t_{\rm dd}$,  both in units of $\lambda$.
  }
  \label{Fig:current}
\end{figure}
\begin{figure}[t]
  \includegraphics[width=0.48\textwidth]{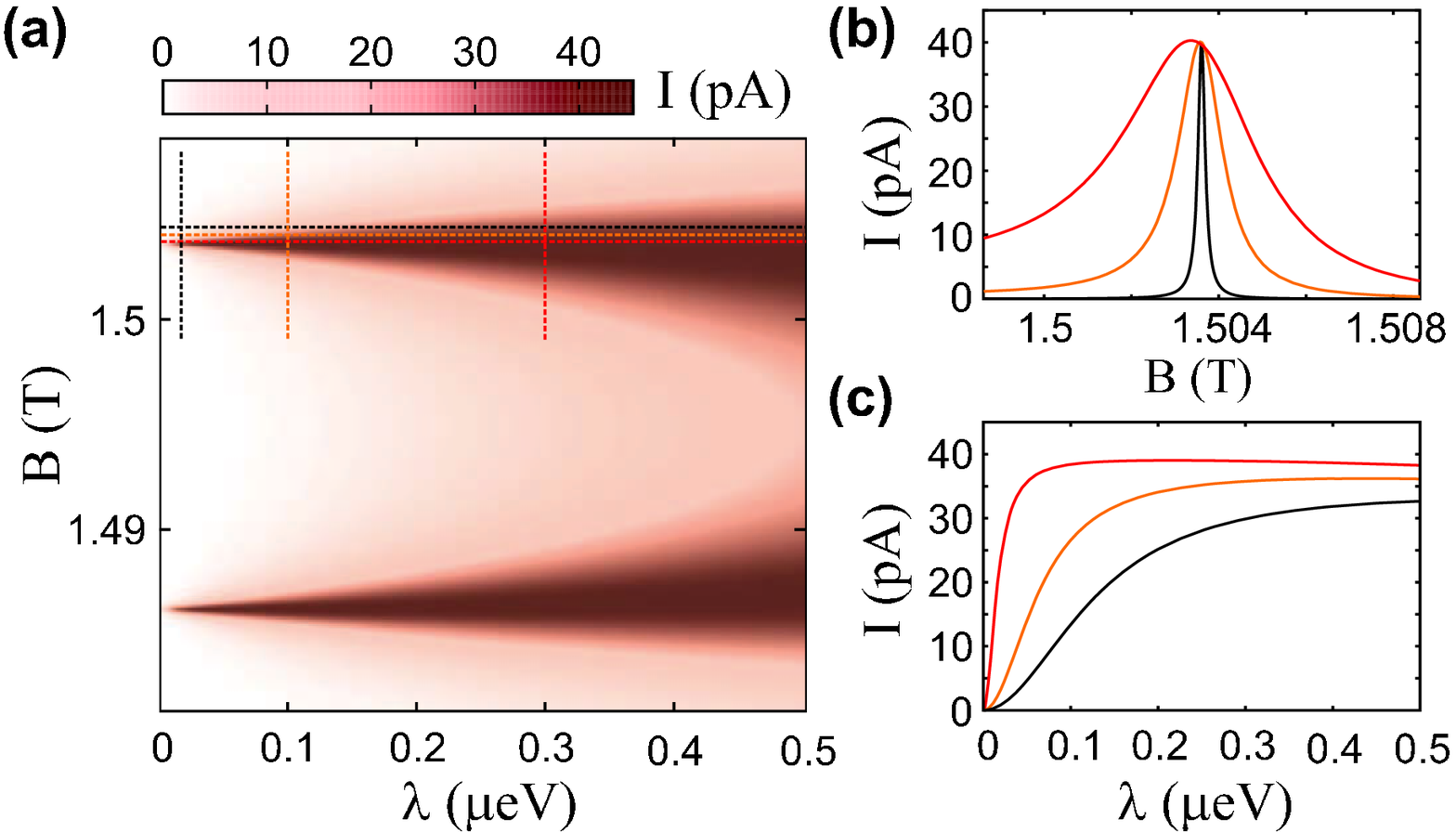}
  \caption{
    (a) Current as function of the magnetic field $B$ and the coupling $\lambda$.
    In (b) and (c) cuts in the parameter space are shown, as indicated in (a).
    With the typical parameters~\cite{Note2}
    and $t_{\rm dd}=\unit[1.0]{\mu eV}$, $\Gamma = \unit[1]{GHz}$,
    we estimate the maximal leakage current $\sim \unit[40]{pA}$.}
  \label{Fig:current-lambda}
\end{figure}
\emph{Readout using Pauli spin blockade.}
We suggest to detect the vibration directly by measuring the charge current in a double dot setup in the spin-blockade regime, as shown in \Fig{Fig:setup}(b). The dots are coupled with interdot tunnel amplitude $t_{\rm dd}$, controlled by the center gate voltage, and by Coulomb charging effects. Only the left dot experiences the time-dependent spin-orbit field due to spin-phonon coupling to the vibrations. In the high magnetic field regime, we can focus on those energy levels of the left and right quantum dot, $\alpha\in\{\mathrm{L,R}\}$ differing only by the spin degree of freedom, $s\in\{\uparrow,\downarrow\}$, that we denote by $\epsilon_{\alpha s}$.
We account for both the intradot Coulomb interaction ($U$) and the interdot interaction ($U'$). We tune the double quantum dot into the Pauli spin-blockade regime by gating the right dot such that it is always occupied by an electron with $s=\downarrow$, see \Fig{Fig:setup}(b).
By a proper choice of the bias window one can restrict the transport to involve only the one-- and two-electron states, $\ket{0,\downarrow}, \ket{\uparrow, \downarrow}, \ket{\downarrow,\downarrow}$ and $\ket{0,\uparrow\downarrow}$. Here the first and second index denote the occupation of the $K$ states of the left and right quantum dot, respectively.
The time-dependence of the spin-orbit field induced by the vibrations results in a constant vibration-induced spin-flip amplitude 
$\bracket{\uparrow,\downarrow}{\downarrow,\downarrow}{\op{H}_{\rm s-ph}}={\lambda}/{4}$. 
This has been found by performing a unitary transformation, $\op{U}=\exp( -\i \omega t \,\op{S}_z )$, to the rotating frame and applying the RWA. When tuning with the gate voltages the singlet states $\ket{\uparrow,\downarrow}$ and $\ket{0,\uparrow \downarrow}$ to be resonant, $\epsilon_{\rm L \uparrow} + \epsilon_{\rm R \downarrow} + U' = \epsilon_{\rm R \uparrow} + \epsilon_{\rm R \downarrow}+ U$, transfer of electrons with $s=\uparrow$ from the left lead to the right lead proceeds via the processes $\ket{0,\downarrow} \xrightarrow{} \ket{\uparrow,\downarrow} \xrightarrow{t_{\rm dd}} \ket{0,\uparrow \downarrow} \xrightarrow{} \ket{0,\downarrow}$.
However, as soon as an electron with spin $s=\downarrow$ enters the left dot, transport is blocked, until the spin-phonon coupling induces a transition between triplet and singlet states, $\ket{\downarrow,\downarrow} \xrightarrow{\lambda} \ket{\uparrow,\downarrow}$.
This physical picture is confirmed by our calculation of the leakage current $I= W_{(0,\uparrow \downarrow) \rightarrow (0,\downarrow)} P_{(0,\uparrow \downarrow)}$, where $W_{(0,\uparrow \downarrow) \rightarrow (0,\downarrow)}$ is a transition rate and the occupation probability $P_{(0,\uparrow \downarrow)}$ for state $\ket{ 0,\downarrow\uparrow}$, 
is calculated from a master equation, cf. Ref.~\onlinecite{Palyi:kx}, neglecting thermal broadening effects for mK temperatures ($V_{\rm bias},\Delta \gg k_{\rm B}T$).
\Fig{Fig:current} shows the leakage current as function of the two experimentally tunable parameters, $B$ and $t_{\rm dd}$ (through the central gate). It exhibits two sharp resonances where the spin-flip rate becomes maximal; this is reached by tuning the magnetic field such that $\Delta \approx \hbar\omega \pm t_{\rm dd}$. The width of these peaks -- as function of both $B$ and $t_{\rm dd}$ -- is on the order of $\lambda$, see the upper panel of \Fig{Fig:current}.
The vibrational frequency can be determined
\footnote{
Note, that for the readout one has to measure the absolute field strength. 
The variable $\delta B$, used in \Fig{Fig:current} for clarity, is not a measurable quantity, because the value of $B_{\rm res}$ is unknown.
}
 -- independent of $t_{\rm dd}$ -- from $\hbar\omega = \half [\Delta(B_{1}) + \Delta(B_{2})]$
and the required value of $\Dso$ can be obtained independently by measuring the spectrum in \Fig{Fig:setup}(c).\cite{Jespersen:2010fk}
For $\DKK \ll \abs{ \Dso \pm 2 \muorb B} $ this simplifies to $\hbar\omega \approx \half \muS (B_{1} + B_{2}) - \Dso$ and $\Dso$ can be read off directly at $B=0$ in \Fig{Fig:setup}(c)(see arrows).
As shown in \Fig{Fig:current-lambda}(a), the coupling constant $\lambda$ determines the broadening of the leakage current peaks and also slightly their positions for a fixed value of $t_{\rm dd}$. Therefore, the readout depends strongly on the coupling constant $\lambda$ and through this on the deflection amplitude, $\frac{\mathrm{d} u_x}{\mathrm{d}z} = \lambda / \Dso$. It determines the resolution of the magnetic field, required  to make the effect, enabling frequency readout, measurable, see \Fig{Fig:current-lambda}(b). In \Fig{Fig:current-lambda}(c), we also show the current as a function of the coupling strength, being proportional to the vibration amplitude.
The sensitivity of the near-to resonant current to the vibration amplitude can be tuned with the magnetic field, which renders also the amplitude readout possible, by comparing different slightly off-resonant current values with calculations. In the limit of ground state fluctuations ($\frac{\mathrm{d}u_x}{\mathrm{d}z} = 9 \times 10^{-5}$ and $\lambda \approx\unit[0.016]{\mu eV}$), the broadening of the current is given by $ B\approx \unit[0.2]{mT}$. In conclusion, this indicates, that our spin-based readout scheme may be even able to resolve vibrational ground state fluctuations when a magnetic field resolution in the regime of mT is available.

We acknowledge funding by the JARA Seed Fund, the DFG (FOR 912) and the Ministry of Innovation NRW.

\emph{Note:}
While completing the manuscript we became aware of a complementary work
exploring the spin-phonon coupling in a CNT in the quantum limit. \cite{Palyi11}

\bibliography{references}

\end{document}